\documentclass[pdflatex,sn-mathphys]{sn-jnl}

\usepackage{color,soul,xcolor}
\DeclareUnicodeCharacter{2061}{}
\newcommand{\fGLMMMethodName}{\emph{dMEGA }}
\newcommand{\denote}{\triangleq}

\jyear{2022}%

\theoremstyle{thmstyleone}%
\theoremstyle{thmstyletwo}%

\theoremstyle{thmstylethree}%
\usepackage{amssymb}

\newcommand{\comment}[1]{}

\begin{document}

\title[Article Title]{Federated Generalized Linear Mixed Models for Collaborative Genome-wide Association Studies}


\author[1]{\fnm{Wentao} \sur{Li}}\email{wentao.li@uth.tmc.edu}

\author[1,2]{\fnm{Han} \sur{Chen}}\email{Han.Chen.2@uth.tmc.edu}

\author*[1]{\fnm{Xiaoqian} \sur{Jiang}}\email{xiaoqian.jiang@uth.tmc.edu}

\author*[1]{\fnm{Arif} \sur{Harmanci}}\email{arif.harmanci@uth.tmc.edu}

\affil[1]{\orgdiv{School of Biomedical Informatics}, \orgname{University of Texas Health Science Center}, \orgaddress{\city{Houston}, \postcode{77030}, \state{TX}}}
\affil[2]{\orgdiv{School of Public Health}, \orgname{University of Texas Health Science Center}, \orgaddress{\city{Houston}, \postcode{77030}, \state{TX}}}



\abstract{As the sequencing costs are decreasing, there is great incentive to perform large scale association studies to increase power of detecting new variants. Federated association testing among different institutions is a viable solution for increasing sample sizes by sharing the intermediate testing statistics that are aggregated by a central server. There are, however, standing challenges to performing federated association testing. Association tests are known to be confounded by numerous factors such as population stratification, which can be especially important in multiancestral studies and in admixed populations among different sites. Furthermore, disease etiology should be considered via flexible models to avoid biases in the significance of the genetic effect. A rising challenge for performing large scale association studies is the privacy of participants and related ethical concerns of stigmatization and marginalization. Here, we present \fGLMMMethodName, a flexible and efficient method for performing federated generalized linear mixed model based association testing among multiple sites while underlying genotype and phenotype data are not explicitly shared. \fGLMMMethodName first utilizes a reference projection to estimate population-based covariates  without sharing genotype dataset among sites. Next, \fGLMMMethodName uses Laplacian approximation for the parameter likelihoods and decomposes parameter estimation into efficient local-gradient updates among sites. We use simulated and real datasets to demonstrate the accuracy and efficiency of \fGLMMMethodName. Overall, \fGLMMMethodName's formulation is flexible to integrate fixed and random effects in a federated setting. 
}

\keywords{Genome-wide Association Testing, Population Stratification, Federated Learning}



\maketitle
\section{Introduction}
Genome-wide association studies (GWAS) are dominant methods for discovering genetic variants that explain the genetic component of phenotypic variance. As the sequencing costs are decreasing, there is great incentive to perform large-scale association studies to increase the power of the studies~\cite{26690481, 21867570}. Currently, population-scale joint genotyping and phenotyping efforts such as AllofUs and UKBiobank generate very large resources that provide great opportunities for extensive analysis of genotype-phenotype relationships~\cite{25826379, 31412182, 15199942}. In addition, there are other efforts that aim at focusing on certain phenotypes such as TOPMed~\cite{33568819}, ADSP~\cite{29055816}, TCGA~\cite{25691825}, and GTEx~\cite{23715323}.

There are a number of standing challenges around performing large scale GWAS in existing datasets. 
Association tests are confounded by numerous factors such as population stratification, which can be especially important in multiancestral studies and in admixed populations~\cite{16862161}. Most of the multiancestral studies are performed as meta-analyses~\cite{29531354, 30698716} and may make it more challenging to correct biases compared to a pooled individual-level data analysis among sites in a collaborative GWAS setting~\cite{23724904,24719363}. Furthermore, binary and continuous traits should be modeled using appropriate models to avoid biases in the significance of the genetic effect. It has been shown previously that binary traits are more appropriately analyzed using generalized linear models (GLM) compared to linear models because generalized models can naturally represent the categorical/binary nature of case/control study designs~\cite{27018471, prentice1979logistic}. In addition, there can be complex relationships among samples (such as cryptic relatedness), which makes it necessary to account for random polygenic effects that may otherwise bias association signals. Furthermore, increasing sample sizes requires extensive collaboration among large institutions, but data sharing (among institutions) may be restricted under diverse regulations such as HIPAA~\cite{31182664} and GDPR~\cite{29395455}. Consequently, a rising concern for performing large-scale collaborative association studies is the consideration of privacy and related ethical concerns of stigmatization and marginalization~\cite{32601475, 35246669}. 

Although there are increased incentives around sharing data and making discoveries,  regulations are enacted on legislative level for stricter protection of personal genetic data from open sharing. This creates a major hurdle for international collaborations. The most basic data protection is performed by lengthy data transfer agreements that authorize users' access to data repositories (e.g. dbGAP~\cite{24297256} and European EGA~\cite{34791407}). The agreements only establish accountability and do not meaningfully protect data, as data is still stored and analyzed in plaintext. On technical domain, differential-privacy~\cite{10.1007/11787006_1}, homomorphic encryption~\cite{10.1145/1536414.1536440}, and secure multiparty computation~\cite{lindell2020secure} enable provably privacy-aware data analysis. Differentially-private methods~\cite{26691928, 26525346} are based on noisy data release mechanisms and substantially degrade genetic data utility. Homomorphic encryption (HE)~\cite{34464590, 32398369, 32693823, 34635645} based approaches enable analysis of encrypted data without decrypting it. Although HE-based methods have made orders of magnitude improvement in terms of performance in the last decade, they still require large computational resources. Similarly, secure multiparty computation (SMC) methods~\cite{29734293} rely on the separation of data among multiple entities such that it cannot be recovered by any of the non-colluding entities. SMC-based methods have high data transfer requirements and may not be practically feasible.  

Federated association testing (rooted from Federated Learning approaches~\cite{29500022, 33204939}) among different sites present a viable solution for increasing sample sizes while underlying genotype and phenotype data is not explicitly shared. In federated association testing methods, the association testing is reformulated as an iterative algorithm. At each iteration, each collaborating site computes intermediary statistics using local genotype and phenotype data and the statistics from other sites. Next, the intermediary statistics are shared among the sites with a central server that is aggregated and re-shared to all sites. Federated testing is advantageous from a privacy perspective because the genotype and phenotype data never leave local sites. This way, all sites make use of the pooled individual-level data that would be otherwise isolated in distributed repositories across institutions. 

Here, we present \fGLMMMethodName, a federated generalized linear mixed model that enables federated genetic association testing among collaborating sites.  First, each site utilizes a reference projection-based approach, wherein the genotype data at the site is projected on an existing public genotype panel (e.g., The 1000 Genomes Project) and population-based covariates are computed based on the projected coordinates. Usage of projection is advantageous because it decreases computational requirements by circumventing computation of principal component analysis (PCA) among the sites and minimally impacts accuracy. In addition, the computation of population-level covariates does not require data to be pooled and does not incur privacy risks. Next, \fGLMMMethodName performs federated association testing using the fixed (such as population covariates) and random effects. In this step, the sites locally calculate intermediate statistics that are sent to a central server, which aggregates the statistics from all sites and shares them with all sites. After a number of iterations, the algorithm converges and final results are calculated. We demonstrate the accuracy and efficiency of \fGLMMMethodName using simulated and real datasets.


\section{Results}\label{sec_Results}

\subsection{Overview of \fGLMMMethodName}\label{subsec_Overview}
\begin{figure}[h]
    \centering
    \includegraphics[scale=0.3]{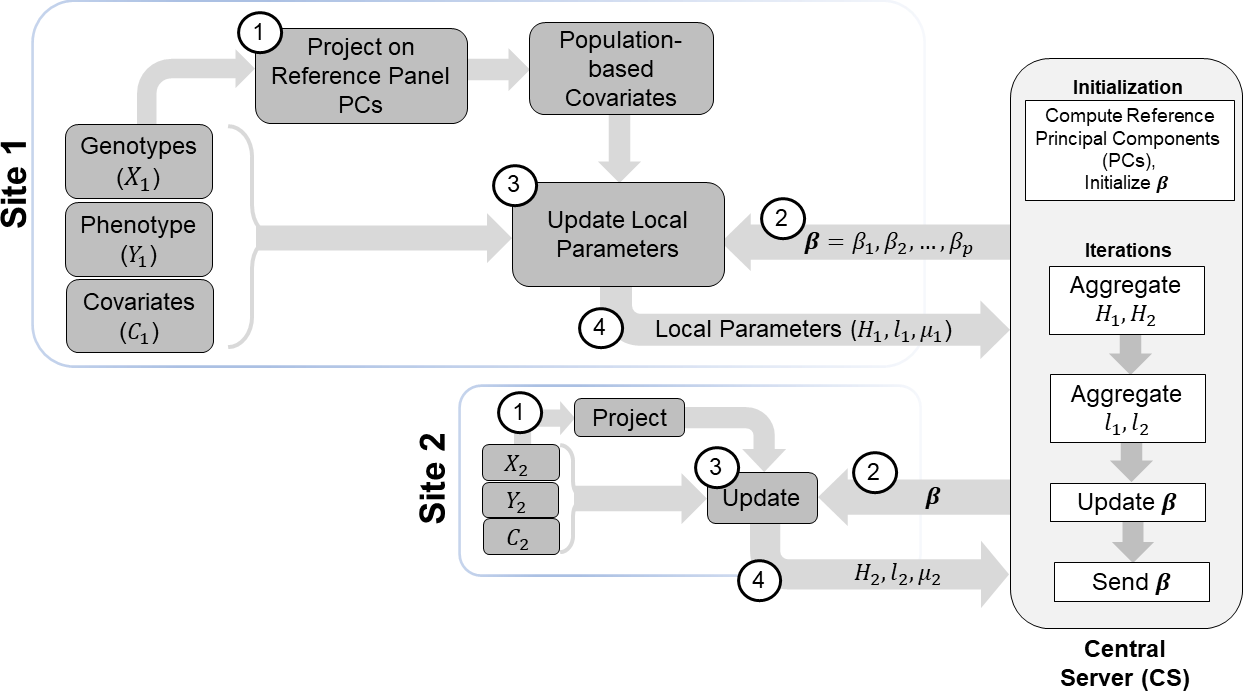}
    \caption{Illustration of federated association testing workflow for two sites named Site-1 and Site-2. Each site holds genotype, phenotype, and covariate datasets. Each site first downloads the reference panel principal components (PCs) and projects the genotypes to generate the population-based covariates (Step 1). Next, the initial parameters are downloaded from the central server (CS) (Step 2). Using the local genotype, phenotype, and merged covariate data, each site updates the local parameters (Step 3) and sends them to CS (Step 4). After receiving the local parameters from both sites, CS aggregates the parameters and sends the updates parameters to all sites. Steps 2, 3, and 4 are performed until the model converges. Step 1 is performed only once at the before iterations.}
    \label{Fig:Overview}
\end{figure}

Fig.~\ref{Fig:Overview} shows the steps of federated association testing workflow. First, the sites project their genotype data on the principal components computed from a reference panel. The reference panel dataset represents a comprehensive population-based information pool. The projected coordinates are used as population-based covariates (fixed effect). Next, each site computes the local testing statistics (gradients, effect sizes, Hessian matrices) and send them to the Central Server (CS). At each site, the likelihood is approximated by Laplace approximation and gradients are calculated using the local data and the current parameters (Methods). The Central Server collects intermediate model statistics during the federated learning process, aggregates the site-specific parameters to compute the global model parameters, and sends the parameters to the sites for the next iteration. The individual-level data (genotypes, phenotypes, and covariates) is not shared with other sites or the central server in the inference.

\subsection{Projection-based Population Stratification}
We first tested whether \fGLMMMethodName's the projection-based population stratification can be used for performing population structure correction in the context of a simple linear model. This approach enables a large decrease in computation cost by circumventing the need for a full PCA of the genotype data pooled from all sites\cite{29993695, 29734293, 32398369, 32132732} and relying on much simpler computation of projections in population-based covariate computation that is used for population-stratification. 

\begin{figure}[h]
    \centering
    \includegraphics[scale=1.5]{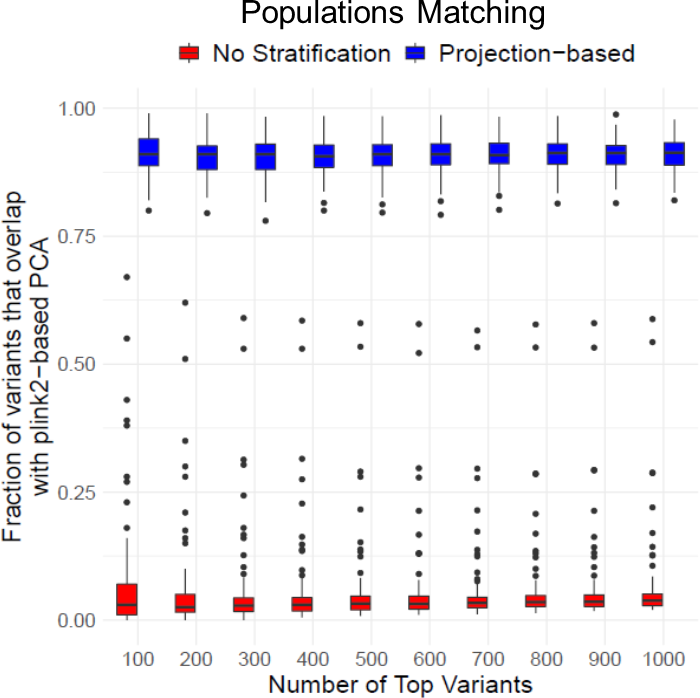}(a)
    \includegraphics[scale=1.5]{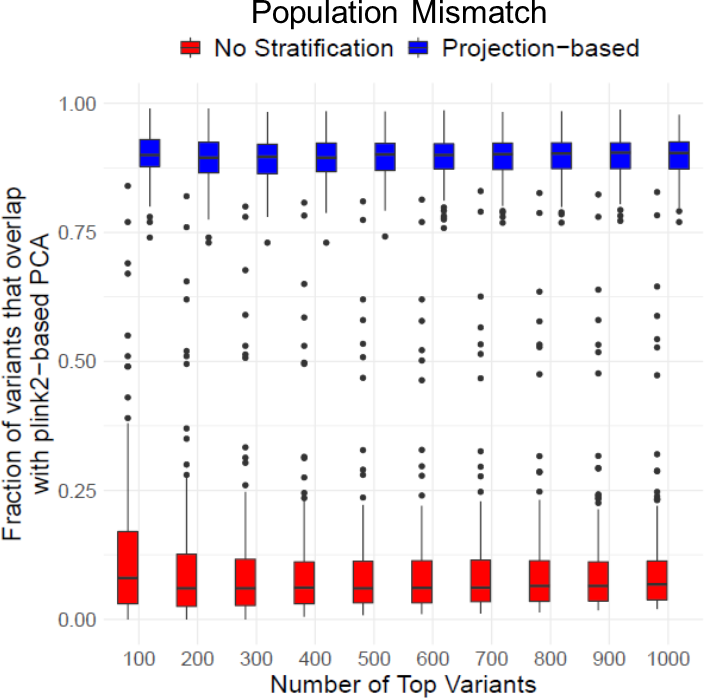}(b)
    \caption{Comparison of most significant variant concordance between Projection-based population stratification and PCA-based stratification among 100 simulated GWAS. (a) The comparison of significant variant concordance for matching population panels. X-axis shows the number of top variants. Y-axis shows the concordance fraction. Blue boxplots depict the concordance between projection-based stratification and PCA-based stratification. Red boxplots show the concordance between GWAS with no population stratification and GWAS with PCA-based stratification. (b) Concordance of most significant variants when projection is performed with a mismatching set of reference populations.}
    \label{fig:GWAS_sims}
\end{figure}

We first simulated 100 GWAS studies where 20 variants with allele frequency below 0.1 were randomly selected as causal with effect on phenotype. We also assigned population and gender-specific biases on the phenotype to introduce population and gender-specific effect. For each simulated study, we simulated genotypes for 3000 individuals with a corresponding quantitative phenotype that is computed regarding genotypes and covariates. We finally ran plink2 in three configurations to perform GWAS with and without population stratification: (1) We ran plink2 with its default PCA to generate population-based covariates; (2) We ran plink2 with covariates generated by projection-based covariate computation using top 6 PCs; (3) We ran plink2 without population stratification as a control to ensure that population stratification is indeed necessary in  GWAS. As a first test scenario, we used CEU, MXL, YRI populations from The 1000 Genomes Project for simulating genotypes and used the same populations as reference to compute the projection-based population covariates, i.e, the projection and simulation ancestries are exactly matched. Overall, p-values and effect sizes from GWAS with projection-based population correction match fairly well to default PCA-based population stratification in plink2 (Fig \ref{suppfig:GWAS_sims}a, b). In comparison, the GWAS without population correction gives fairly discordant and biased results (Fig \ref{suppfig:GWAS_sims}c,d). We next used the GIH, CHB, PEL populations as the reference panel populations to test for mismatches in the simulated and reference populations. We observed that similar results held where projection-based population correction yields good concordance with plink2’s default PCA-based population correction (Fig \ref{suppfig:GWAS_sims}e,f). GWAS p-values and effect sizes without population correction yields fairly discordant results when compared to default correction (Fig \ref{suppfig:GWAS_sims} g, h). We also observed that top variants detected from projection-based correction matches accurately to default correction (Fig \ref{fig:GWAS_sims} a, b). Overall, these results show that projection-based population stratification can be effective for correction of population-specific biases to a large extent. Most importantly, this approach can be implemented efficiently in a secure domain with much better overall performance compared to a full PCA-based population correction. We utilize projection-based population stratification to estimate the population covariates in the GWAS analysis.

\subsection{Accuracy Comparison with Centralized Association Tests}
First, we compared the accuracy of \fGLMMMethodName in collaborative setting by comparing the association results with the centralized model as computed by lme4\cite{Volodina2022-xi}. We used the genotypes and phenotypes data from the database of genotypes and phenotypes (dbGaP) with accession number phg000049 that comprises 3,007 individuals (1,266 case, 1,279 controls, 462 unknown). We ran \fGLMMMethodName after partitioning data into 3 sites based on the k-means algorithm on the genotypes data and used the projection-based covariates for population stratification. The sites were treated as the additive random intercept effect, and this can be more efficient than treating them as fix-effects when the number of sites is exploding. In the comparisons, we focused on the top 10,000 SNPs that were reported by plink~\cite{17701901} version 2 as most significantly associated with the disease status. We also evaluated the utility of projection-based population stratification and correction by estimating the population-based covariates using the 2,504 individuals in the 1000 Genomes Project (Methods). 

We first compared the top SNPs at different significance levels using two stratification approaches to evaluate their effect, which are shown in Table \ref{table:comps} using projections on top 4 and 6 components as covariates in population stratification. Overall, both methods exhibit high concordance with the centralized model (lme4). 

\begin{figure}[h]
    \centering
    \includegraphics[scale=0.4]{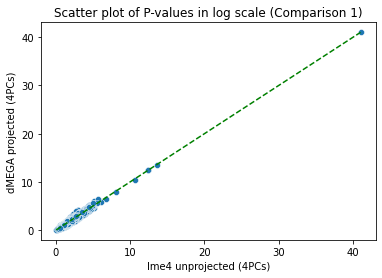}\textbf{a}
    \includegraphics[scale=0.4]{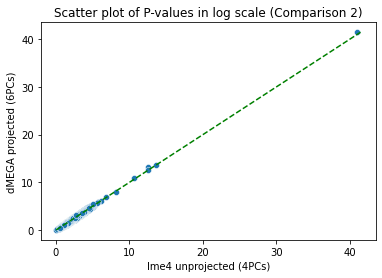}\textbf{b}
    \caption{\textbf{a}, Scatter plot of p-values from comparison 1; \textbf{b}, Scatter plot of p-values from comparison 2.}
    \label{fig:scatter}
\end{figure}

We next pooled all of the SNPs and plotted the assigned p-values, which is shown in Fig.~\ref{fig:scatter}. Consistent with previous result, we observed that using 6 components exhibit a higher concordance of significance levels (Spearman correlation $\rho=0.99$ for 6-components vs $0.97$ for 4-components.)

\begin{figure}[h]
    \centering
    \includegraphics[scale=0.4]{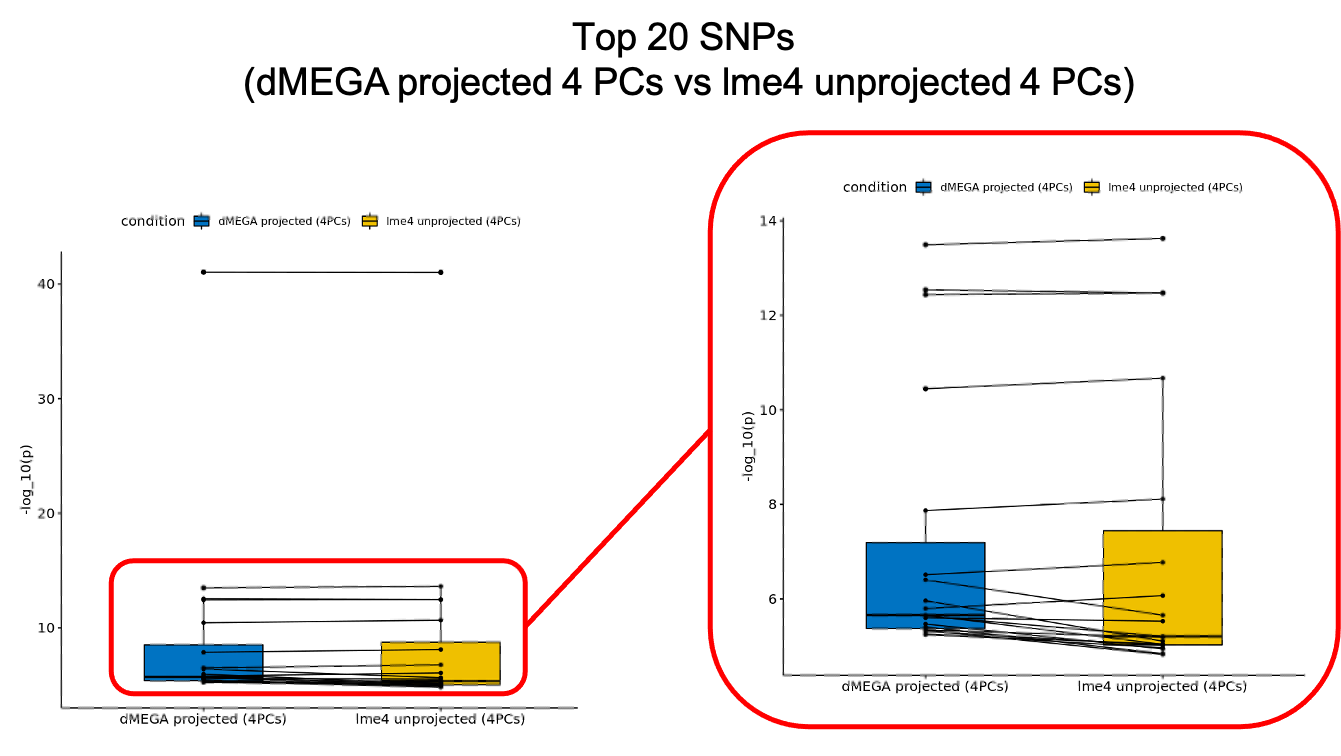}\textbf{a}
    \includegraphics[scale=0.4]{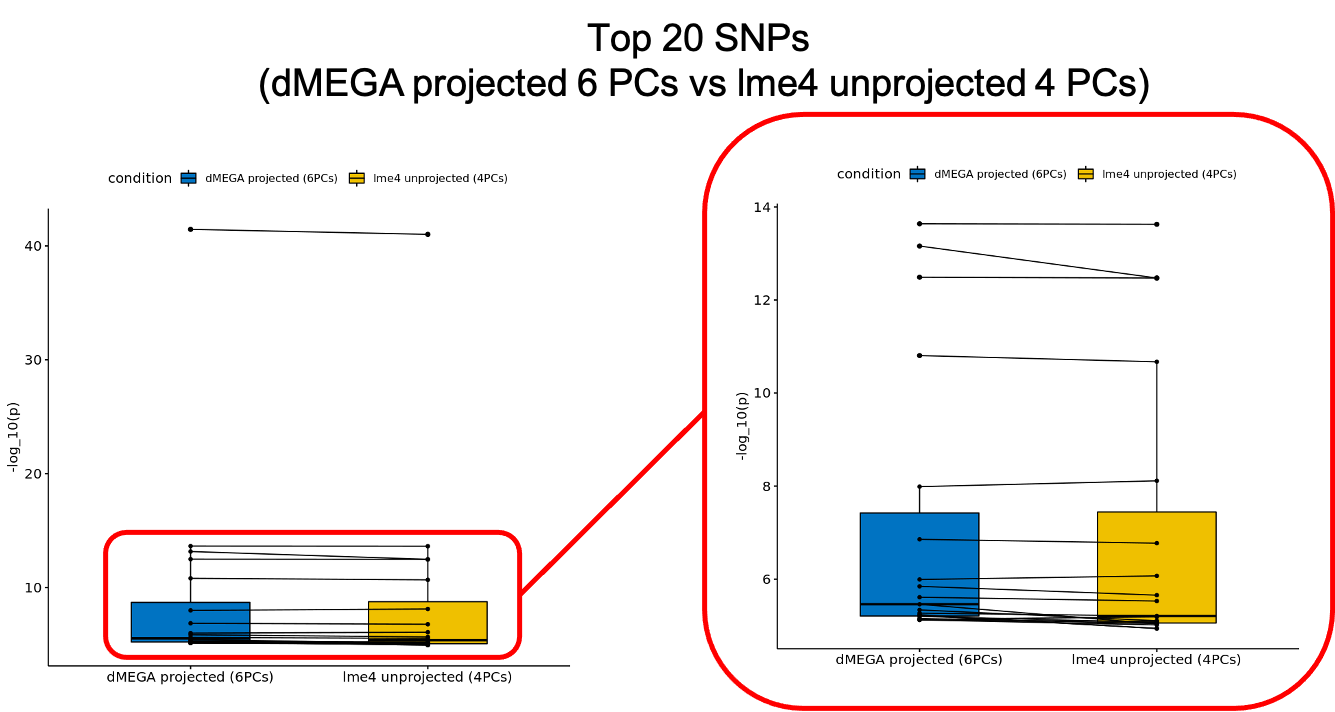}\textbf{b}
    \caption{\textbf{a}, Paired boxplot of comparison 1; \textbf{b}, Paired boxplot of comparison 2}
    \label{fig:boxplot}
\end{figure}

\begin{figure}[h]
    \centering
    \includegraphics[scale=0.33]{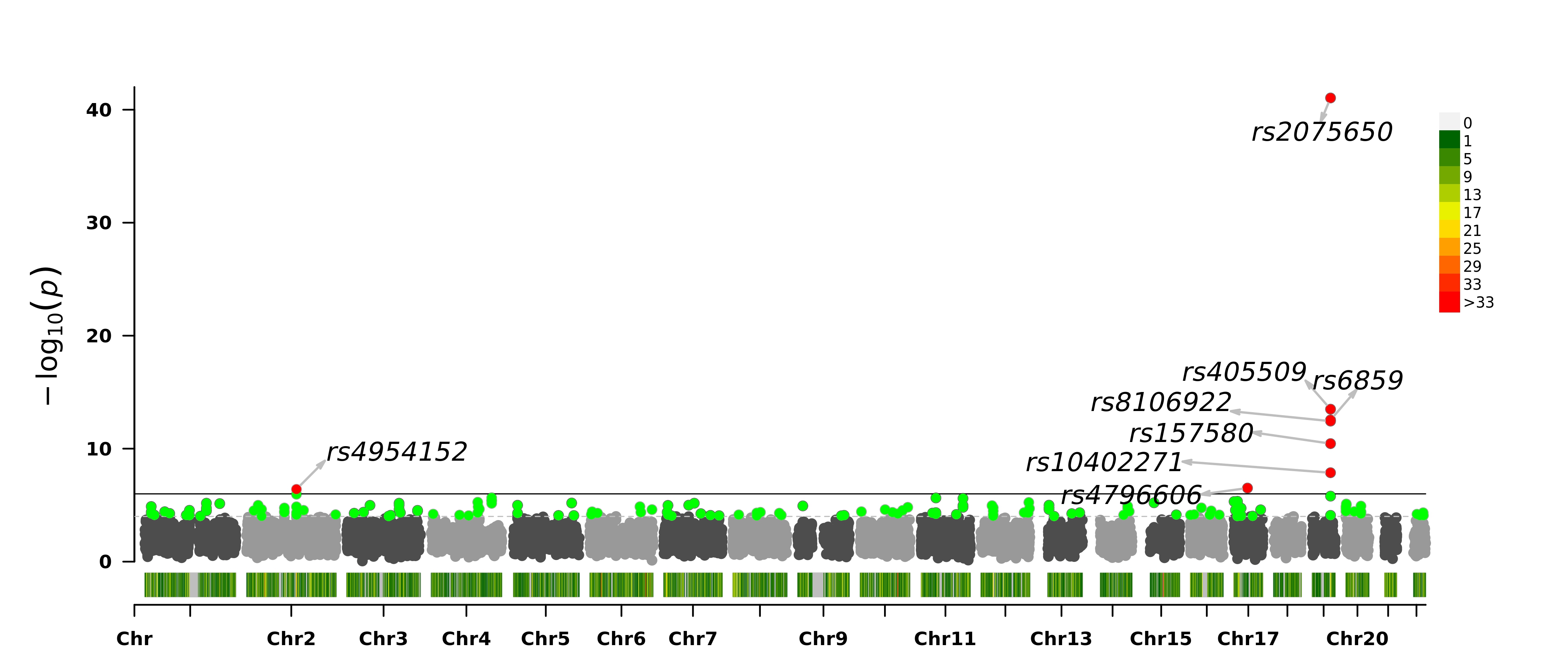}
    \caption{Manhattan plot: \fGLMMMethodName with projected datasets on 4 PCs}
    \label{fig:man_fGLMM_projected_4PCs}
\end{figure}

\begin{figure}[h]
    \centering
    \includegraphics[scale=0.33]{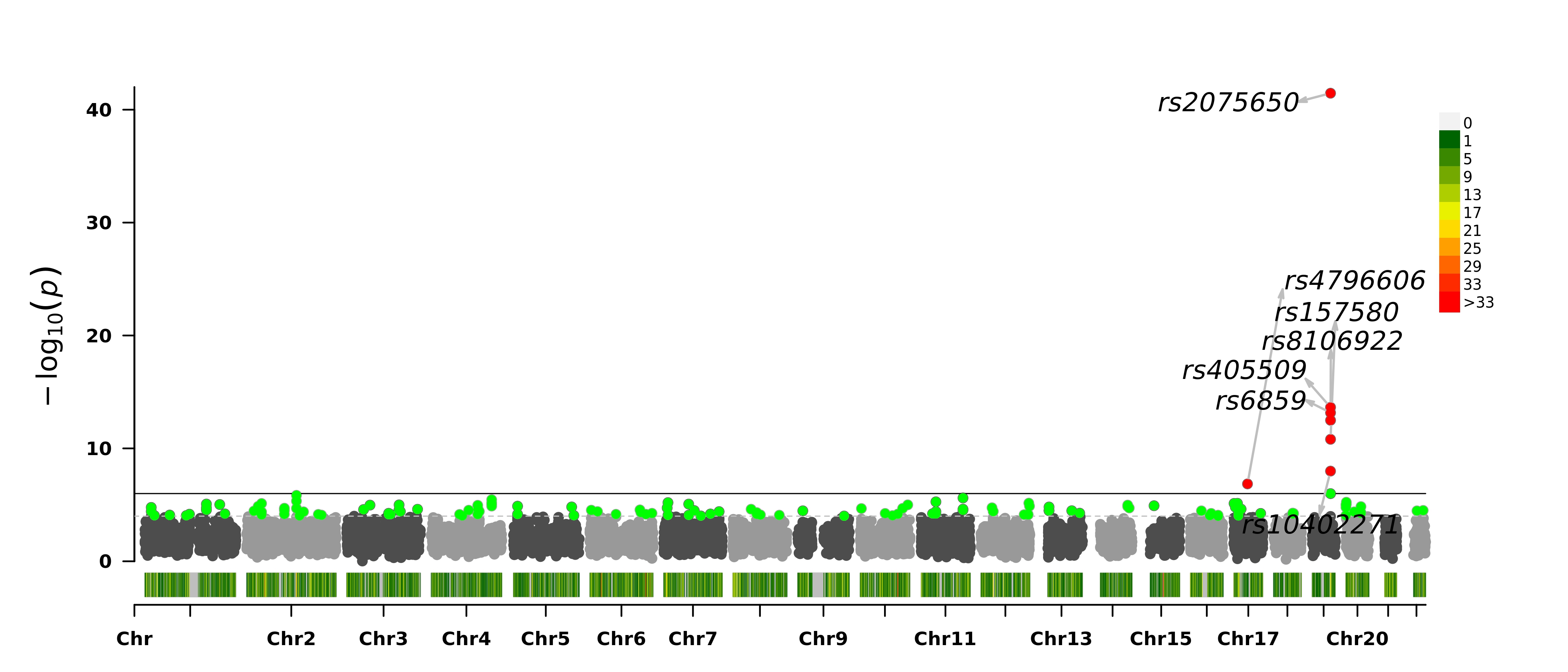}
    \caption{Manhattan plot: \fGLMMMethodName with projected datasets on 6 PCs}
    \label{fig:man_fGLMM_projected_6PCs}
\end{figure}

\begin{figure}[h]
    \centering
    \includegraphics[scale=0.33]{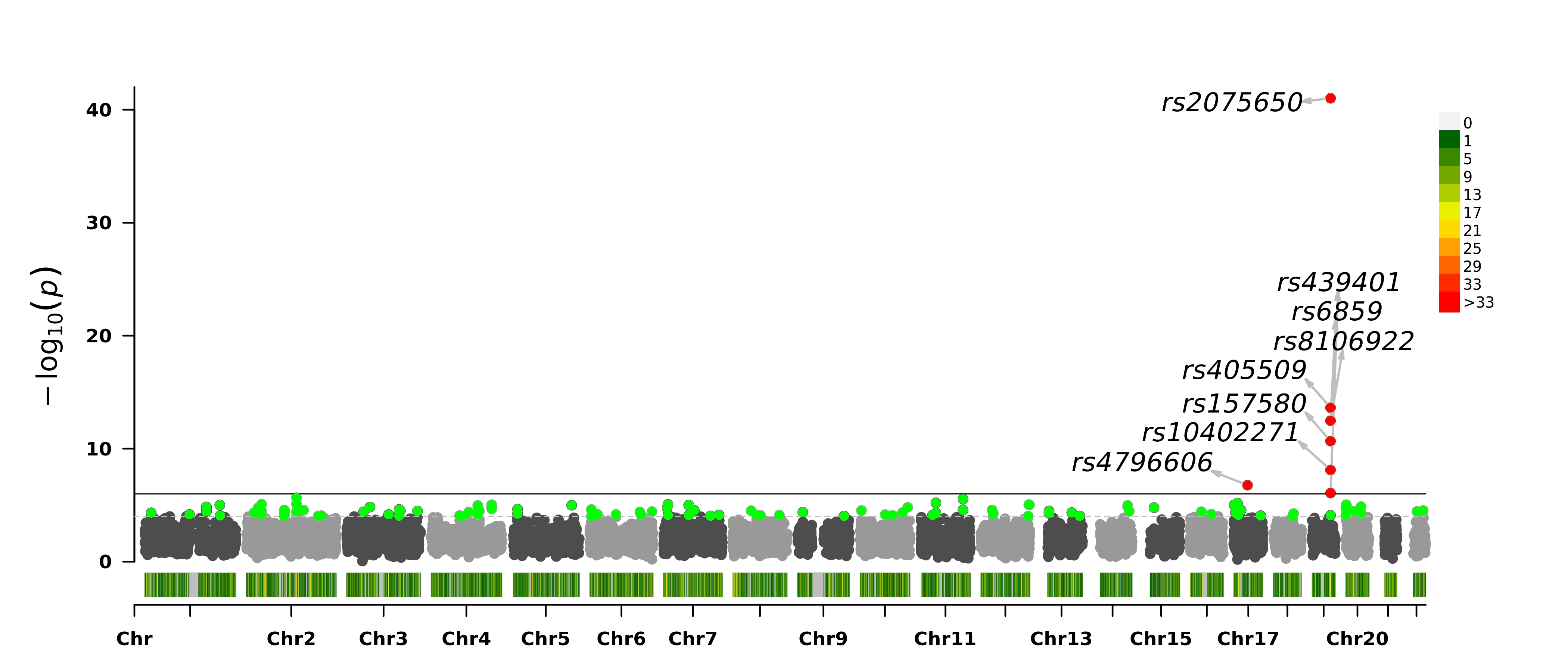}
    \caption{Manhattan plot: lme4 with unprojected datasets on 4 PCs}
    \label{fig:man_lme4_unprojected_4PCs}
\end{figure}

We next compared the ranks assigned to most significant SNPs by the two approaches when they are compared to the centralized model (Fig.~\ref{fig:boxplot}). Overall, there is fairly high concordance in the top SNPs and their rankings. Qualitatively, we observed ranking consistency to the centralized model is higher for population correction using 6-component projection compared to 4-component projections. Table~\ref{table:sig} We finally visually evaluated the genome-wide distribution of the SNP significance, assigned by the 2 projection approaches and the centralized model Fig.~\ref{fig:man_fGLMM_projected_4PCs},\ref{fig:man_fGLMM_projected_6PCs},\ref{fig:man_lme4_unprojected_4PCs}. As expected, all methods find the most significant associations on chromosome 19 with high concordance.


\begin{table}[h]
\centering
\caption{Performance of predicting significant SNPs under various significant level $\alpha$}
\begin{tabular}{ccccc}
\hline\hline
         & \multicolumn{4}{l}{Comparison 1: \fGLMMMethodName on projected 4 PCs vs `lme4' on unprojected 4 PCs}   \\ \hline
$\alpha$ & precision    & recall      & F1-score   & Significant SNPs    \\
$10^{-5}$ & 0.580645     & 0.818182    & 0.679245   & 22                  \\
$10^{-6}$ & 0.875        & 0.875       & 0.875      & 8                   \\
$10^{-7}$ & 1            & 1           & 1          & 6                   \\ \hline\hline
         & \multicolumn{4}{l}{Comparison 2: \fGLMMMethodName on projected 6 PCs vs `lme4' on unprojected 4 PCs}   \\ \hline
$\alpha$ & precision     & recall      & F1-score   & Significant SNPs    \\
$10^{-5}$ & 0.68         & 1          & 0.81        & 19                  \\
$10^{-6}$ & 1            & 0.88        & 0.93       & 8                   \\
$10^{-7}$ & 1            & 1           & 1          & 6                   \\ \hline\hline
\end{tabular}
\label{table:comps}
\end{table}

\begin{table}[h]
\centering
\caption{Correlation statistics in two comparisons.}
\begin{tabular}{rcc}\hline
            & Comparison 1 & Comparison 2 \\\hline
Spearman correlation   & 0.9734       & 0.9909\\
Pearson correlation   & 0.9509        & 0.9845\\
\hline     
\end{tabular}
\label{table:corr}
\end{table}

\begin{table}[h]
\centering
\caption{20 Selected SNPs with high significance.}
\begin{tabular}{lcccc}\hline
SNP        & CHR & \begin{tabular}[c]{@{}c@{}}\fGLMMMethodName \\ projected 6PCs\end{tabular} & \begin{tabular}[c]{@{}c@{}}\fGLMMMethodName \\ projected 4PCs\end{tabular} & \begin{tabular}[c]{@{}c@{}}lme4 \\ unprojected 4PCs\end{tabular} \\\hline
rs2075650  & 19  & 3.48E-42                                                        & 9.14E-42                                                        & 9.64E-42                                                         \\
rs405509   & 19  & 2.30E-14                                                        & 3.22E-14                                                        & 2.36E-14                                                         \\
rs8106922  & 19  & 3.24E-13                                                        & 3.65E-13                                                        & 3.37E-13                                                         \\
rs6859     & 19  & 6.93E-14                                                        & 2.89E-13                                                        & 3.37E-13                                                         \\
rs157580   & 19  & 1.57E-11                                                        & 3.58E-11                                                        & 2.13E-11                                                         \\
rs10402271 & 19  & 1.03E-08                                                        & 1.35E-08                                                        & 7.70E-09                                                         \\
rs4796606  & 17  & 1.39E-07                                                        & 3.07E-07                                                        & 1.69E-07                                                         \\
rs439401   & 19  & 1.02E-06                                                        & 1.60E-06                                                        & 8.50E-07                                                         \\
rs4954152  & 2   & 1.43E-06                                                        & 3.95E-07                                                        & 2.21E-06                                                         \\
rs2507880  & 11  & 2.45E-06                                                        & 2.51E-06                                                        & 2.96E-06                                                         \\
rs2939753  & 11  & 5.38E-06                                                        & 2.30E-06                                                        & 6.19E-06                                                         \\
rs11649731 & 17  & 7.32E-06                                                        & 4.55E-06                                                        & 6.47E-06                                                         \\
rs2924943  & 2   & 4.59E-06                                                        & 1.09E-06                                                        & 7.84E-06                                                         \\
rs7592667  & 2   & 7.55E-06                                                        & 2.38E-05                                                        & 8.10E-06                                                         \\
rs1526528  & 7   & 6.34E-06                                                        & 1.05E-05                                                        & 8.54E-06                                                         \\
rs1798296  & 12  & 7.03E-06                                                        & 5.74E-06                                                        & 9.01E-06                                                         \\
rs1471263  & 4   & 3.47E-06                                                        & 2.20E-06                                                        & 9.18E-06                                                         \\
rs6078239  & 20  & 8.82E-06                                                        & 9.74E-06                                                        & 9.26E-06                                                         \\
rs12320530 & 12  & 1.29E-05                                                        & 1.97E-05                                                        & 9.54E-06                                                         \\
rs7222487  & 17  & 7.52E-06                                                        & 4.82E-06                                                        & 9.63E-06                                                        \\\hline
\end{tabular}
\label{table:sig}
\end{table}


\section{Timing, Memory, and Data Transfer}\label{time_memory_benchmarks}
Our experiment was done in a computation environment of 96 threads (24 Cores) Intel(R) Xeon(R) Platinum 8168 CPU @ 2.70 GHz, 1.5 TB memory, Ubuntu 16.04.7 LTS, Python 3.10.4.
There is negligible difference between running 6PCs and 4PCs in \fGLMMMethodName. Federated computation for SNP will take 20 seconds and 200 MB memory in average to compute. The total communication cost of 4PCs and 6 PCs federated model are 80 KB and 125 KB on average. This cost includes, for each iteration, each site transmit 88 Bytes of random effect coefficient, 144 Bytes of computational intermediates, and 72 Bytes of fixed effects coefficients.

\section{Discussion}\label{sec_discussion}
We presented \fGLMMMethodName for federated generalized linear mixed modeling. \fGLMMMethodName is readily applicable to collaborations where data sharing at the summary statistic level can be deployed. Unlike previous methods that rely on a computationally intensive federated PCA for performing population stratification and correction, \fGLMMMethodName makes use of projection on existing reference panels and to correct for population biases. As the size and diversity of existing panels increase, we foresee that projection-based bias correction can prove more accurate. The projection has very small computational requirements and can be performed at each site before federated analysis. 

Another similar method distributed Penalized Quasi-Likelihood (dPQL) \cite{dpql} used penalized quasi-likelihood to approximate the objective function of GLMM. dPQL simplifies the maximization of the log-likelihood function of GLMM to  fitting a linear mixed model. However, dPQL has inherited drawbacks. For example, dPQL depends strongly on the estimated variance components \cite{breslow1993approximate}, but the inference of variance parameters in dPQL may be biased \cite{ju2020laplace} due to its linearity assumption. \fGLMMMethodName approximates the marginal distributions, and it provides more robust estimation since \fGLMMMethodName considers the marginal predicted ratios associated with each local site, not just identifying samples with particularly high predicted ratios\cite{breslow1993approximate}. 

\fGLMMMethodName has several limitations that warrant further research. While our federated testing approach has small network traffic requirements, each local site is required to handle high computational load. This is a general challenge among federated learning methods. Considering that the gene association tests may involve millions of variants along with large number of phenotypes, the center server aggregation layer can be outsourced to cloud whereby, while the data are kept locally.  

From privacy perspective, \fGLMMMethodName shares only summary statistics between the central server and the sites. In that regard, it is necessary that the central server is a trusted entity (such as NIH) and that all sites are expected to execute \fGLMMMethodName in an honest manner. As summary statistics may leak information, honest-but-curious entities can perform re-identification attacks. This is, however, a general concern in federated learning frameworks and not specific to \fGLMMMethodName. Our approach does not pose a direct risk to the reference panel because projection requires only the principal components. Thus, restricted reference panels from underrepresented populations can be utilized in these computations. It is still worth noting that the principal components are types of summary statistics and can leak information that may be used to re-identify participation using previously described attacks~\cite{18769715}.


To increase the confidentiality of \fGLMMMethodName,  we can utilize noise addition to hide local intermediate information, denoted as $I_i$ (i.e. local sample size, local gradient, local Hessian, local mixed effects, and local standard error), during communication. This idea has been developed in the HyFed\cite{nasirigerdeh2021hyfed} framework, which introduces a server called \textit{Compensator} to collect the local noise $N_i$ values from each client and send the aggregated noise, i.e., $N=\sum_iN_i$, to the CS. 
In this process, each client generates local noise $N_i$ from a Gaussian distribution with zero mean and a variance of $\sigma^2$. Then, each client will mask the intermediate statistics $I_i$ using the noise $N_i$, to generate $I'_i=I_i+N_i$, and send the noisy statistic to CS. Simultaneously, each site sends the noise levels to the \textit{Compensator}. When all clients finish their communication, CS unmasks the global information of interest $I=\sum_i I_i=\sum_i I'_i-N$ via deducting the aggregated noise $N$ provided by \textit{Compensator}.

\section{Conclusion}\label{sec_conclusion}
\fGLMMMethodName is a new method for performing federated generalized linear mixed model (GLMM) fitting for association studies. The general framework is flexible and can be used to optimize GLMMs in a federated manner on other types of data. \fGLMMMethodName makes use of a reference projection-based population stratification to correct for the biases introduced by ancestral differences between the subjects. Projection-based stratification is advantageous because it helps circumvent performing a computationally expensive federated principal component analysis among the institutions. \fGLMMMethodName provides good performance by utilizing the site-wise mixed-effects. We believe \fGLMMMethodName has the potential to impact the collaborative GWAS community by addressing the challenges in data isolation, cohort's bias, and federated computation costs.

\section{Methods}\label{sec_Methods}
The goal of \fGLMMMethodName is to detect significant SNPs that are associated with specific diseases or phenotypes in a federated manner. In our assumption, genotype and phenotype data are stored cohort-wise throughout several entities (e.g., research institutions or hospitals). Each entity is presumed to be prohibited from sending original data. By constructing a logistic regression model with mixed effects, data holders will update the global model with local information bias considered. Notice that the communication process does not put data at risk due to \fGLMMMethodName will only ask data holders for model information, such as gradients.

\subsection{Projection-based Calculation of Population Covariates}
\fGLMMMethodName first centers the genotype matrix for each individual and projects the samples on a  reference panel that is shared among the sites. In the context of privacy-aware analysis, this is a reasonable assumption because the sites can make use of numerous publicly available panels. For \fGLMMMethodName, we use The 1000 Genomes Project panel that comprises 26 diverse sets of populations that are geographically sampled over the world.

The reference panel is first processed at the central server. This is done by performing principal component analysis (PCA) on the reference panel by decomposition of the genotype covariance matrix, i.e., $P \cdot P^T = \Pi \cdot \Lambda \cdot \Pi^T$, where $\Pi_{N, S}$ denotes the full set of principal components of reference panel genotype matrix $P_{N, S}$ for $N$ genetic variants and $S$ samples in the reference panel, where $S=2,504$ for The 1000 Genomes Project population data. We use $\kappa$ top principal components (columns) of this matrix in our projection step.

After the reference panel is processed by the central server, the principal components are sent to collaborating sites. It should be noted that the reference panel is processed once at the central site at the beginning of the computations. The central server does not share the reference panel genotypes directly with the sites. The components do not represent direct risk to the reference panel individuals. This is advantageous for utilizing the restricted population panels, such as the ToPMED panel~\cite{33568819}. 

\begin{equation}
    \tilde{G}_{i,j} = G_{i,j} - \frac{1}{N} \cdot \sum_k P_{i, k}
\end{equation}
where $\tilde{G}$ denotes the centered genotype matrix. 

\begin{equation}
    c_{k,j} = \sum_i \tilde{G}_{i,j} \cdot \Pi_{i,k}, k<\kappa
\end{equation}
where $c_{k,j}$ denotes the $k^{th}$ covariate for $j^{th}$ individual. 

\subsection{Projection-based Covariate Computation with The 1000 Genomes Sample}
In our experiments, we used The 1000 Genomes Project's phase 3 genotypes as the reference panel available at \href{http://ftp.1000genomes.ebi.ac.uk/vol1/ftp/release/20130502/}. We used the bi-allelic SNPs and subsampled the variants to utilize 77,531 variants. We generated the top 4 and 6 principal components for the 3,007 individuals in the genotype dataset.

\subsection{Federated association test}
We introduce a federated association test algorithm based on Generalized Linear Mixed Model. Assume that there will be $k$ institutions that hold genotype and phenotype data, and that each institution's database consists of $n_i$ subjects. Let the total number of patients be denoted by $n=\sum_i^k n_{i=1}$. Here, we consider site-wise mixed-effects, denoting $\mu_i$ as the mixed-effect of institution $i$, as well as shared fixed-effects $\beta$. The genotype dataset at institution $i$ denoted as $X_i$ (Matrix of $N$ variants and $n_k$ individuals), and phenotypes denoted as $Y_i$ (vector of length $n_k$).
Thus, the mixed model of each site can be represented as
\begin{align*}
\mathbb E[Y_i\mid\mu_i, X_i]&=g^{-1}(X_i\beta+\mu_i)\\
\mu_i&\sim\mathcal N(0,\sigma)
\end{align*}
where $g^{-1}(\cdot)$ is the inverse of the link function (i.e., a logit function for logistic regression/binary traits) that defines the relationship between the linear combination of the predictors (genotypes, covariates, and random effects) to the mean of the phenotype. Here, we focus on the random intercept effect at site $i$, $\mu_i$, which follows a normal distribution with mean $0$ and variance $\sigma$. In this scenario, $\mu_i$ is constant for individuals on the same site. Across the sites, $\mu_i$ is normally distributed across sites.

For a binary trait (i.e. case/control study), the conditional probability distribution of the phenotype given the variant genotypes and covariates can be written as
\[
P(Y_{ij} = 1\mid X_{ij})=\int_{\mu_i} g^{-1}(X_i\beta+\mu_i)\phi(\mu_i)d\mu_i
\]
where $\phi$ denotes the probability density function for normal distribution with mean 0 and hyperparameter variance $\sigma$. Thus, the likelihood function of the joint distribution can be formulated as
\[
\mathcal L(\beta,\sigma)=\prod_{i=1}^k\int_{-\infty}^{+\infty}\prod_{j=1}^{n_i} P(\beta,\mu_i\mid X_{ij},Y_{ij})P(\mu_i\mid\sigma)d\mu_i
\]


The optimization of the likelihood function is a non-tractable problem because the integral over the random effects does not have a closed form representation. we utilize Laplace approximation and derive the approximated objective function 

\[
 l(\beta,\sigma)\denote\log{\mathcal L(\beta,\sigma)}=\sum_{i=1}^k\log\mathcal L_i(\beta,\hat\mu_i)\denote\sum_{i=1}^k l_i(\beta,\hat\mu_i)
\]


Hence, the goal is to optimize the approximated objective function above. Compared to the centralized (all data pooled in one repository) inference, the optimization in federated learning settings is based on iterations of (1) Calculation of the intermediate statistics computed using each institution's local data and (2) aggregation of the statistics by a central server (CS). We describe the steps in more detail below:

\textbf{Initialization.}The federated learning will start with a central server $CS$ that connects to $k$ distributed local data repositories. Initial modeling information requests will send to each participant $P_i$.
\begin{itemize}
    \item Number of PCs $p$
    \item A list of SNPs' name $S$
    \item A list of sample size across participants $N$
\end{itemize}

\textbf{Step 1.} $CS$ will initiate model parameters $\beta_{(0)}$ for each distributed model with SNP in list $S$. And each local repository $P_i$ computes model's intermediates and send back to $CS$

\textbf{Step 2.} $CS$ updates model's parameter $\beta_{new}$ with aggregated information from global gradients $l'(\mu)=\sum_il'_i(\mu_i)$, global hessian $H{(\mu)}=\sum_iH_i(\mu_i)$, and previous fixed-effects $\beta_{prev}=(\beta_1, \dots, \beta_k)$. The update is done by Newton's method $\beta_{new}=\beta_{prev}-l'/H$. Then send $\beta_{new}$ to each $P_i$.

\textbf{Step 3.} Each $P_i$ will follow \textbf{Step 2} until model is converged with criteria $\Delta\beta$ and $\Delta\mu$ below threshold $10^{-6}$.

\textbf{Step 4.} The $CS$ will compute the local standard errors
from $P_i$, then return inference statistics (e.g. Z score, P-values). 

All the information in communication is summarized in table below
\begin{table}[h]
\centering
\begin{tabular}{ll}
\hline
\multicolumn{1}{c}{$P_i$ to $CS$} & \multicolumn{1}{c}{$CS$ to $P_i$} \\ \hline
 Number of PCs $p$: \textbf{scalar}  & Current working SNP's name $S$: \textbf{character}  \\
 A list of SNPs' name $S$: \textbf{list} & Global gradient $l$: $p+1$ \textbf{vector}  \\
 Sample size $n_i$: \textbf{scalar} &  Global Hessian $H$: $(p+1)\times (p+1)$ \textbf{matrix}\\
 Local gradient $l_i'$: $p+1$ \textbf{vector}  & Global parameter $\beta$: $p+1$ \textbf{vector}  \\
 Local Hessian $H_i$: $(p+1)\times (p+1)$ \textbf{matrix}  & Inference statistics: $p+1$ \textbf{vector} \\ 
 Local mixed effect $\mu_i$: \textbf{scalar} & \\
 Local Standard Error $SE_i$: $p+1$ \textbf{vector} & \\\hline
\end{tabular}
\end{table}

\section{Data Sources and Experimental Setup}
We used genotype-phenotype data obtained from database of Genotypes and Phenotypes (dbGaP) with accession number phs000168 for our experiments available for General Research Use (GRU). This dataset contains $575,003$ variants genotyped by Illumina Human610-Quad version 1 platform over $3,007$ individuals. Raw data is processed and formatted with plink2~\cite{17701901}. The alternate alleles reported by the array platform were re-coded using in-house scripts to ensure that they were concordant with The 1000 Genomes Project. Any variant for which we could not resolve by strand were excluded. We next used plink2's "--glm" option to calculate the baseline association signals. We next filtered the SNPs and identified the SNPs with top $10,000$ variants with the strongest association signal to the phenotype. 

The reference panel is obtained from the 1000 Genomes Project FTP portal at \href{ftp://ftp.1000genomes.ebi.ac.uk/vol1/ftp/}{ftp://ftp.1000genomes.ebi.ac.uk/vol1/ftp/}. We processed 1000 Genomes dataset by first  excluding the SNPs with minor allele frequency (MAF) smaller than $5\%$. We next overlapped the variants with the re-coded array variants, which yielded $155,076$ common variants. To decrease computational requirements, we focused on variants on the 22 autosomal chromosomes and further sub-sampled the remaining variants to generate the final $77,315$ variants. These variants were used to generate the principal components and population-based covariates in the projection step.

To evaluate \fGLMMMethodName, we compared it with a baseline method using the linear mixed model implemented in R package `lme4'~\cite{bates2014fitting}. Our  experiments were designed as table below:
\begin{table}[h]
\centering
\begin{tabular}{r|cc}\hline
            & Distributed & Pooled \\\hline
Projected   & \fGLMMMethodName\textsuperscript{*\textdagger}       & R(lme4)\textsuperscript{\textdagger}\\
Unprojected & \---       & R(lme4)\textsuperscript{*}\\\hline     
\end{tabular}
\end{table}


We will focus on two comparisons:
\begin{enumerate}
    \item (Denoted in \textdagger) \fGLMMMethodName in projected and distributed data and baseline in projected and pooled data. 
    
    While the datasets are the same (projected), this comparison aims to show the performance of \fGLMMMethodName in distributed datasets.
    \item (Denoted in $\ast$) \fGLMMMethodName in projected and distributed data and baseline in unprojected and pooled data. 
    
    The datasets are of different between \fGLMMMethodName method (using projected datasets) and baseline method (using unprojected datasets). This comparison will show the capability of projection combining with federated learning.
\end{enumerate}

\backmatter





\section*{Acknowledgments}

XJ is CPRIT Scholar in Cancer Research (RR180012), and he was supported in part by Christopher Sarofim Family Professorship, UT Stars award, UTHealth startup, the National Institute of Health (NIH) under award number R01AG066749 and U01TR002062.
\newpage
\pagebreak
\bibliography{manuscript.bib}


\newpage

\begin{appendices}

\section{Figures}

\begin{figure}[h]
    \centering
    \includegraphics[scale=1.5]{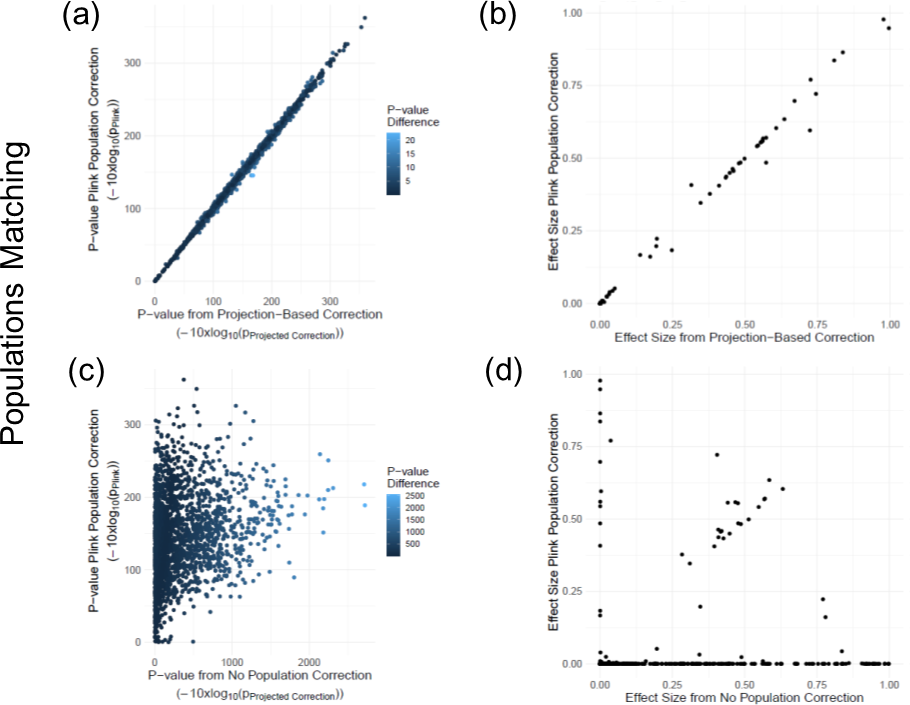}
    \includegraphics[scale=1.5]{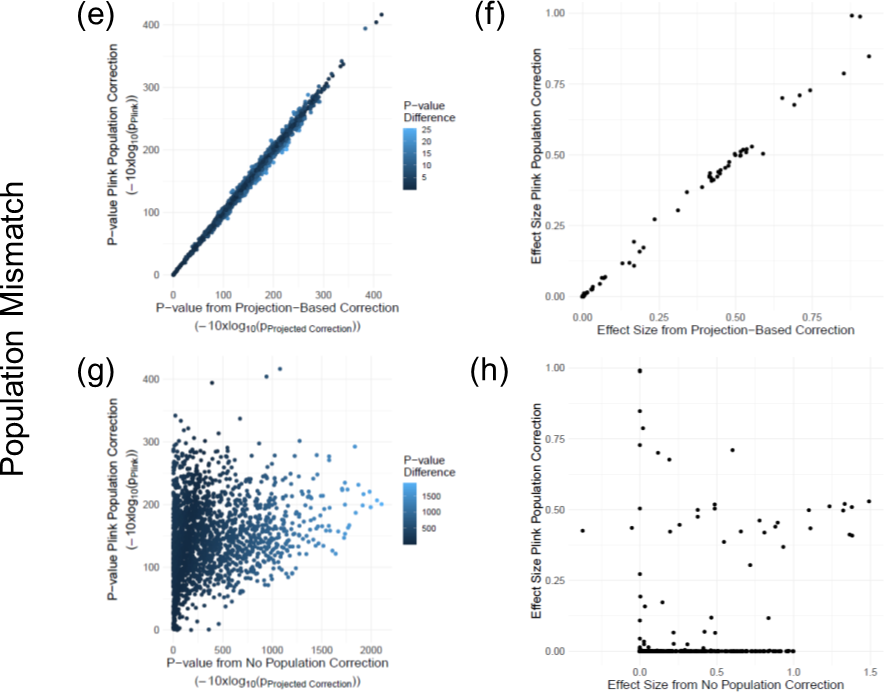}
    \caption{Comparison of projection-based population stratification with PCA-based stratification among 100 simulated GWAS studies. (a) Scatter plot shows the p-value $(-10×log_10⁡(p-value))$ estimates with plink2’s PCA-based population correction (y-axis) versus p-value estimates from population correction using projection-based estimation of population covariates. (b) Scatter plot shows the effect size estimates with plink2’s PCA-based population correction (y-axis) versus effect size estimates from population correction using projection-based estimation of population covariates. (c) Scatter plot shows the p-value estimates with plink2’s PCA-based population correction (y-axis) versus p-value estimates without population correction. (d) Scatter plot shows the effect size estimates with plink2’s PCA-based population correction (y-axis) versus effect size estimates without population correction. (e) Scatter plot shows the p-value $(-10×log_10⁡(p-value))$ estimates with plink2’s PCA-based population correction (y-axis) versus p-value estimates from population correction using projection-based estimation of population covariates where the reference population is not matching to the populations of GWAS individuals. (f) Scatter plot shows the effect size estimates with plink2’s PCA-based population correction (y-axis) versus effect size estimates from population correction using projection-based estimation of population covariates for non-matching reference population. (g) Scatter plot shows the p-value estimates with plink2’s PCA-based population correction (y-axis) versus p-value estimates without population correction for non-matching reference population. (h) Scatter plot shows the effect size estimates with plink2’s PCA-based population correction (y-axis) versus effect size estimates without population correction for non-matching reference population.}
    \label{suppfig:GWAS_sims}
\end{figure}

\begin{figure}[h]
    \centering
    \includegraphics[scale=0.3]{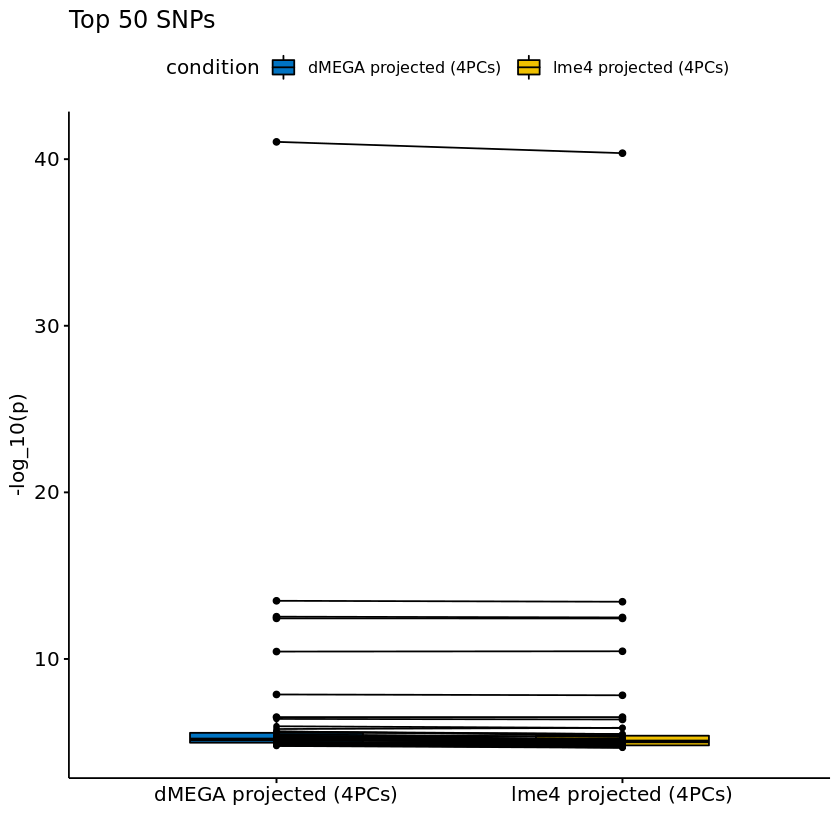}
    \includegraphics[scale=0.3]{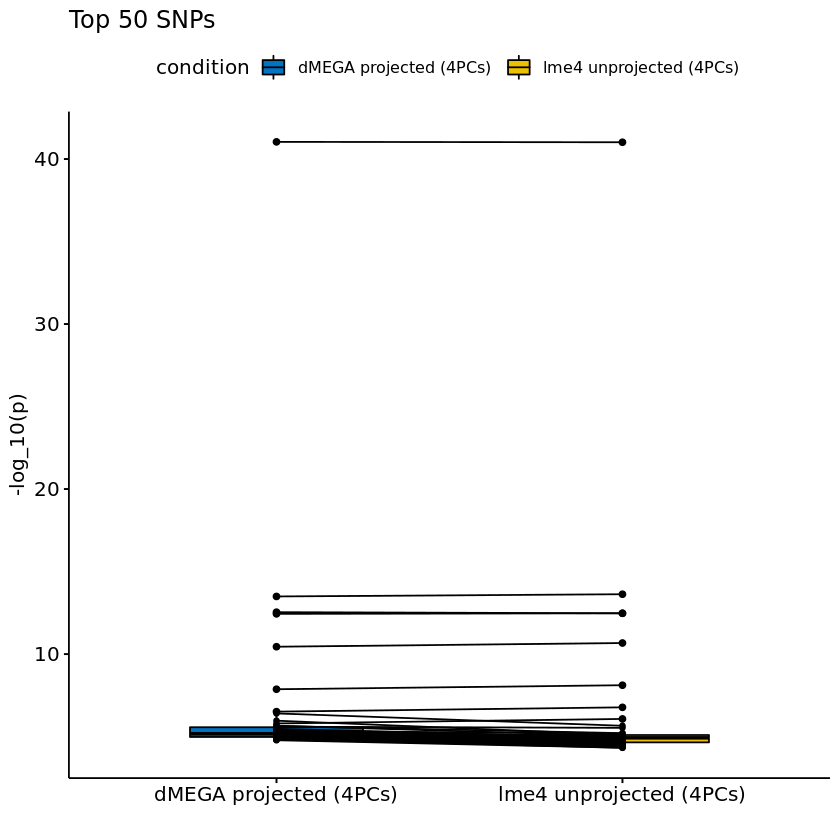}
    \caption{(Left) Paired boxplot of comparison \textdagger; (Right) Paired boxplot of comparison $\ast$}
    \label{fig:top50}
\end{figure}

\begin{figure}[h]
    \centering
    \includegraphics[scale=0.3]{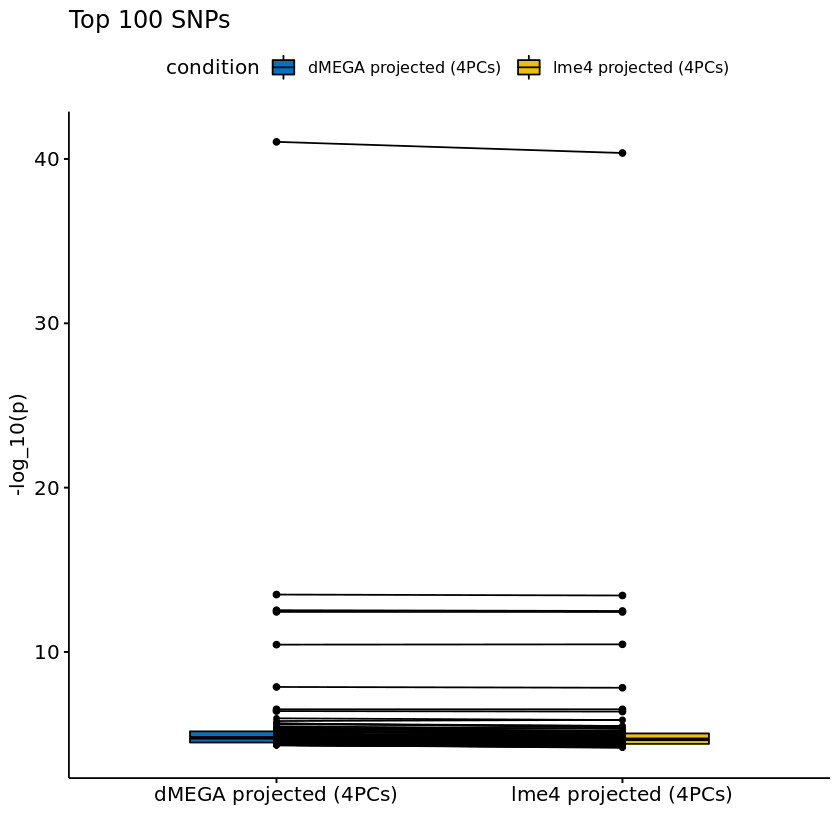}
    \includegraphics[scale=0.3]{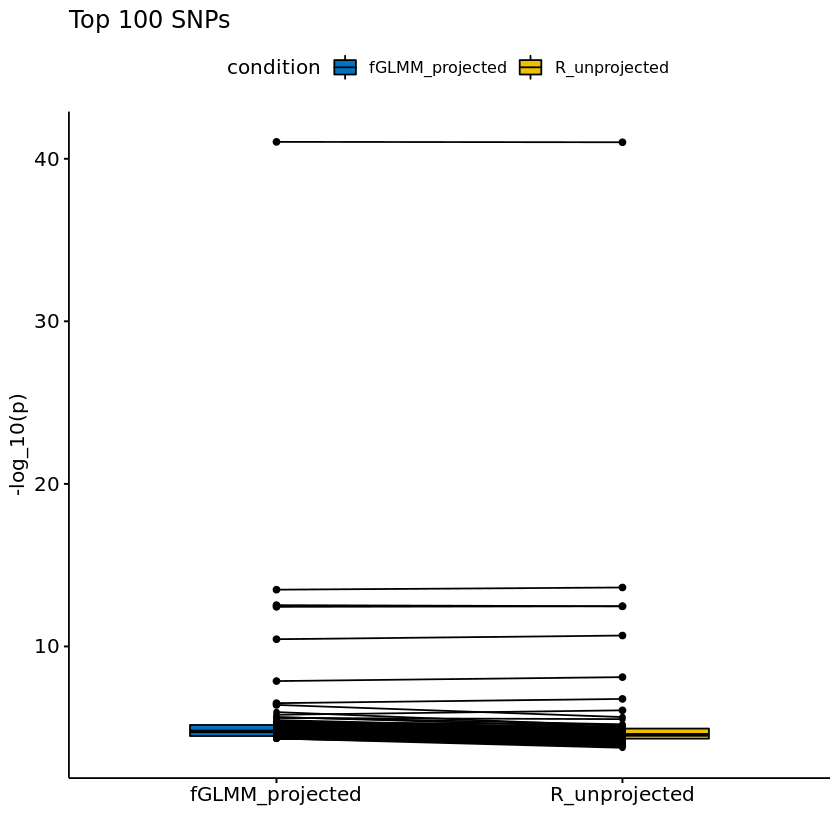}
    \caption{(Left) Paired boxplot of comparison \textdagger; (Right) Paired boxplot of comparison $\ast$}
    \label{fig:top100}
\end{figure}

\begin{figure}[h]
    \centering
    \includegraphics[scale=0.3]{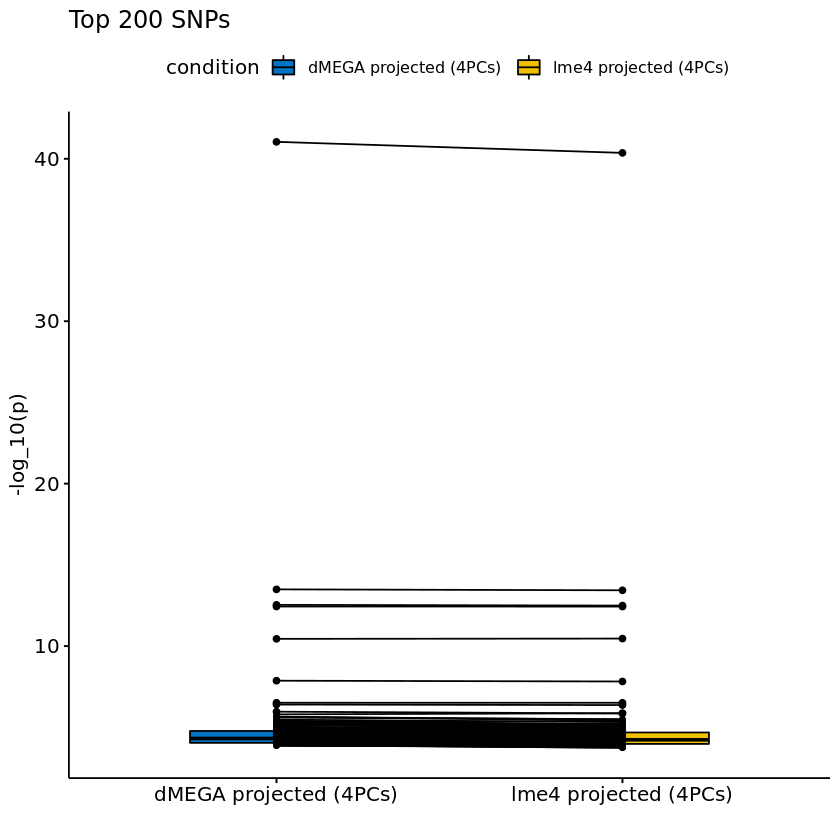}
    \includegraphics[scale=0.3]{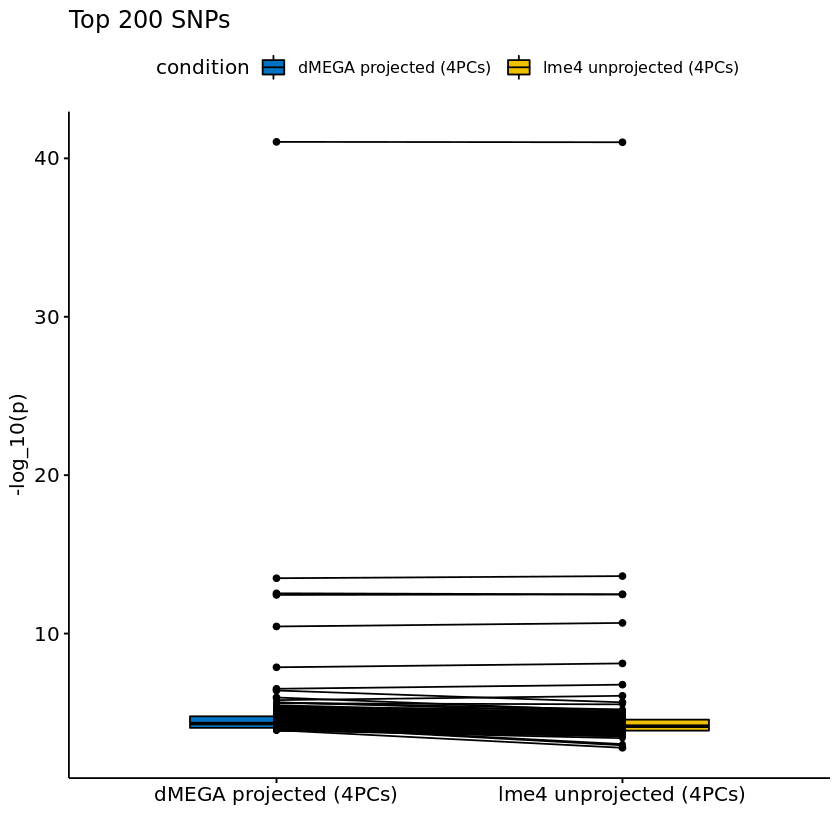}
    \caption{(Left) Paired boxplot of comparison \textdagger; (Right) Paired boxplot of comparison $\ast$}
    \label{fig:top200}
\end{figure}




\end{appendices}

\end{document}